\documentclass[journal]{IEEEtran}
\usepackage{xcolor,soul,framed}
\colorlet{shadecolor}{yellow}
\usepackage{graphicx}
\graphicspath{{../pdf/}{../jpeg/}}
\DeclareGraphicsExtensions{.pdf,.jpeg,.png}

\usepackage[cmex10]{amsmath}
\usepackage{mdwmath}
\usepackage{mdwtab}
\usepackage{eqparbox}
\usepackage{url}
\usepackage{amssymb}
\usepackage{tabularx}
\usepackage{booktabs}
\usepackage{caption}
\usepackage{algorithm}
\usepackage{algpseudocode}
\usepackage{subcaption}
\usepackage{geometry}
\usepackage{array}
\usepackage{float}
\usepackage{cite}
\begin{document}

\title{Security-enhanced Blockchain with Twin-Field Quantum Key Distribution: A Physical Layer enabled Architecture}
\author{Xuan~Li, Yun Mao and~Ying~Guo
\thanks{X. Li and Y. Mao are with the School of Computer Science, Beijing University of Posts and Telecommunications, Beijing 100876, People’s Republic of China.}
\thanks{Y. Guo is  with the School of Computer Science, Beijing University of Posts and Telecommunications, Beijing 100876, People’s Republic of China,  and the Hefei National Laboratory, Hefei 230088, People’s Republic of China.}
}

\maketitle

\begin{abstract}
Quantum computing provides a feasible multi-layered security challenge 
to classical blockchain networks. Quantum blockchains that rely on quantum key distribution (QKD) to establish secure channels can address this feasible threat. Whereas, there are still architecture limitations to practical security resulted in the measurement devices while implementing the QKD-secured blockchains in physical layer. This paper presents a quantum-classical hybrid architecture in a distributed blockchain to address the connectivity and distance limitations of the blockchain-embedded quantum networks. 
A decoupled architecture is designed felicitously so that it pairs a linearly scalable measurement-device-independent (MDI) physical layer with a decentralized consensus.
It can optimize the complexity of infrastructure from quadratic to linear scaling, ascribed to leveraging the twin-field (TF) QKD protocol with the MDI-structurized star topology. Additionally, the dual-key stratification strategy transforms symmetric information-theoretic security into publicly auditable forward-secret blockchain evidence. This architecture can integrate the exact information-theoretic security (ITS) with distributed consensus mechanisms, allowing the scalable system to overcome the potential rate-loss limits inherent in classical security-weakened blockchains.  
\end{abstract}


\IEEEpeerreviewmaketitle

\section{Introduction}
Blockchain establishes a decentralized trustworthy distributed ledger system by integrating a peer-to-peer network, a consensus mechanism and a robust cryptographic framework~\cite{nakamoto2008bitcoin}. The security and integrity of this kind of framework are fundamentally underpinned by classical cryptographic primitives, among which the asymmetric encryption provides secure identity authentication. 

However, quantum computing may present a multi-layered security challenge to classical blockchains~\cite{castiglione2024integrating,fedorov2018quantum,olushola2025cybersecurity}. For example, quantum computers running Shor's algorithm~\cite{Shor} can solve integer factorization and discrete logarithm problems in polynomial time, potentially breaking mainstream asymmetric encryption schemes, such as the elliptic curve digital signature algorithm~\cite{RN3} (ECDSA) and the Rivest-Shamir-Adleman algorithm (RSA)~\cite{RN4}. Accordingly, it may compromise digital signatures and user authentication at the cryptographic layer. Meanwhile, popular computational mechanisms like Proof-of-Work (PoW) are also vulnerable at the consensus layer. Simultaneously, a quantum adversary who performs Grover's algorithm~\cite{Grover} could achieve a quadratic speedup in hash computations. This would enable them to monopolize network hashing power and launch a 51\% attack, subverting the ledger's integrity~\cite{yang2023survey,gharavi2024post,sezer2025pp}. The dual-threats to both cryptography and consensus highlight the vulnerability of classical blockchains and underscore the urgent need for a holistic quantum-resistant architecture rather than piecemeal fixes.

To address the threats that quantum computing poses to blockchain-secured networks~\cite{wazid2024generic}, two parallel and valuable paradigms have emerged. The first paradigm focuses on post-quantum cryptography~\cite{pqc} (PQC), which represents a monumental achievement in modern cryptography. By deploying advanced mathematical structures such as lattice-based and hash-based constructions, PQC successfully hardens the logical software layer of blockchains against known quantum algorithms. With ongoing global standardization efforts, PQC offers a efficient and scalable solution that relies on deeply vetted computational hardness assumptions, establishing itself as a robust and practical choice for securing broad digital infrastructures.

Complementary to the computational paradigm of PQC, the second approach explores physical-layer enhancements through quantum key distribution (QKD)~\cite{kiktenko2018quantum,yang2024qbma,reddy2025quantum}. While PQC secures the algorithmic layer, QKD provides a distinct cryptographic property: information-theoretic security (ITS). This approach is particularly appealing for specific consortium blockchains that seek to decouple their long-term security from computational limits entirely. The security of QKD is strictly governed by the fundamental postulates of quantum mechanics, specifically the Heisenberg uncertainty principle and the no-cloning theorem. In practice, any eavesdropping attempt on the quantum channel inevitably disturbs the fragile quantum states, allowing legitimate nodes to precisely quantify potential information leakage and distill provably secure symmetric keys. Within a distributed ledger architecture, these continuous symmetric keys can be directly integrated with message authentication codes to secure transactions and consensus voting. Consequently, a QKD-enabled physical layer offers a parallel, physics-driven infrastructure that guarantees the theoretical immutability of distributed networks.

Nevertheless, embedding QKD in large-scale multi-node networks such as blockchains usually encounters operational constraints. As the intrinsic point-to-point structure of conventional QKD systems diverges from the decentralized communication topology required by distributed ledgers, establishing secure quantum channels across geographically dispersed nodes has to address kinds of practical issues, including network scalability, efficient key routing among arbitrary nodes, the maintenance of sufficient key generation rates to meet transaction throughput demands. As consequence, overcoming these implementation barriers is a prerequisite for realizing a practical quantum-secured blockchain.

To counter the threats of quantum computing and achieve information-theoretic authentication without relying on vulnerable digital signatures, initial quantum blockchains utilized the BB84 protocol to distribute symmetric keys~\cite{kiktenko2018quantum,reddy2025quantum}. While these keys enable unconditionally secure message authentication rooted in the no-cloning theorem, the application of BB84 to distributed ledger systems faces practical limitations. Owing to the lack of ideal single-photon sources, a feasible implementation with weak coherent pulses seems vulnerable to photon-number-splitting (PNS) attacks~\cite{gottesman2004security,pns1,brassard2000limitations,makarov2009controlling,lydersen2010hacking}. Besides, the BB84 protocol is inherently ill-suited for scalable networks, as its point-to-point architecture necessitates a quadratic $\mathrm{O}(N^{2})$ increase in quantum links for an $N$-node network~\cite{PLOB}. Additionally, the secret key rate decays exponentially with transmission distance, making it difficult to sustain the massive key consumption required by high-frequency consensus voting.

Fortunately, the twin-field (TF) QKD protocol offers a potential solution that extends the scalability of quantum networks~\cite{tfqkd}. 
Regarding the practical security of the TF-QKD system, it adopts a measurement-device-independent (MDI) structure, whereas the single-photon interference occurs at an untrusted central relay~\cite{mdiqkd}. 
This kind of structure renders the system immune to all detector-side channel attacks, allowing the relay to remain untrusted without compromising the practical security of the physical system. 
Intriguingly, it overcomes the fundamental linear rate-loss limit (Pirandola-Laurenza-Ottaviani-Banchi bound, PLOB) inherent to point-to-point protocols~\cite{PLOB}. 
By enabling the secret key rate to scale with the square root of the channel transmittance, the TF-QKD protocol supports the high-rate key generation over inter-city distances. 
As consequence, this architecture facilitates a scalable star-shaped topology, ascribed to an fact that it can connect multiple users via a single central node, reducing the infrastructure complexity and making it intrinsically suitable for large-scale geographically distributed blockchains.

In the light of capabilities of the TF-QKD protocol, this paper proposes an improved quantum-resistant blockchain architecture for consortium networks. Crucially, previous QKD-integrated blockchains often overlook a fundamental cryptographic gap: in a purely symmetric key environment, simple hash functions cannot independently prove the origin of a transaction to third parties (i.e., they fail to provide public non-repudiation). To resolve this limitation, we propose a dual-key stratification strategy. By employing a vector of one-time Wegman-Carter (WC) message authentication codes (MACs)~\cite{WC} alongside a delayed key disclosure protocol, we successfully transform inherently private symmetric verification into a publicly auditable proof. This mechanism enforces a strict separation between transaction evidence, which is rendered publicly verifiable, and consensus security, which remains permanently secret, thereby ensuring long-term ledger immutability.

Strikingly, we propose a quantum-secured Byzantine fault tolerance (BFT) consensus protocol. Unlike the traditional BFT implementations relying on computationally vulnerable digital signatures~\cite{bft1,bft2}, our scheme employs ITS authentication codes derived from the TF-QKD-based physical layer. It enhances the integrity and resilience of the consensus process against both internal malicious nodes and external quantum adversaries. 
Accordingly, this approach ensures the security of network without computational hardness assumptions, ascribed to the integrated authentication within the permission consensus logic.

To the best of our knowledge, this paper proposes a suitable framework that transforms theoretically secure quantum principles into a scalable and practical distributed ledger. The main contributions of this work are summarized as follows:

    \textbf{1) Integrating blockchain fitted with TF-QKD:} We suggest a  quantum-secured blockchain architecture underpinned by the TF-QKD protocol. Compared to mainstream BB84-based systems, our approach leverages the star-shaped quantum link topology and the capability to overcome the PLOB bound, offering superior scalability, extended transmission distances, and higher key rates that are inherently more suitable for large-scale distributed networks.

    \textbf{2) Dual-key enhanced stratification strategy:} We address a fundamental limitation in existing QKD-integrated blockchains, where the strict reliance on symmetric keys inadvertently compromises the system's non-repudiation and public auditability. By implementing a dual-key stratification mechanism, this architecture successfully reconciles symmetric information-theoretic security with the verifiable evidence requirements of blockchains for the first time.

    \textbf{3) Global evaluation of cross-layer architecture:} Diverging from previous results that primarily simulate isolated key generation rates at the QKD physical layer, we establish a comprehensive cross-layer analytical model. By quantitatively evaluating the physical layer's QKD key supply  alongside the logical consensus layer's blockchain cryptographic key consumption, we provide a holistic supply-demand framework, offering a rigorous reference methodology for the performance analysis of future quantum-classical hybrid systems.

The remainder of this paper is organized as follows. 
Section~\ref{sec:preliminaries} outlines the preliminaries and related work. 
Section~\ref{sec:system_architecture} details the proposed architecture, integrating the TF-QKD logical consensus layer with the quantum-secured BFT consensus mechanism. 
The comprehensive security analysis and performance evaluation are presented in Section~\ref{sec:security_analysis} and Section~\ref{sec:performance}, respectively. 
Finally, Section~\ref{sec:conclusion} concludes the paper.

\section{Preliminaries and Related Work}
\label{sec:preliminaries}
\subsection{Quantum Threats on Blockchain}
\label{subsec:classical_and_threats}

A blockchain functions as a decentralized immutable ledger, structurally relying on a layered architecture comprising data, network, consensus, and application layers. The security is strictly predicated on classical cryptographic primitives, such as asymmetric cryptography for digital signatures and cryptographic hash functions for data integrity and consensus. However, the emergence of quantum computing challenges the rigorous assumptions of computational hardness, involving the integer factorization and discrete logarithm problems that underpin these primitives.

In classical architectures, user identity and transaction non-repudiation are both guaranteed by digital signature schemes. The security of these schemes hinges on the computational infeasibility of deriving a private key from a public key using classical cryptographic algorithms. Accordingly, the consensus layer, predominated by PoW, relies on the pre-image resistance of hash functions~\cite{aggarwal2017quantum}. Consistency of the ledger is ensured by the one-CPU-one-vote mechanism, attributed to computational resources to solve a probabilistic puzzle.

Nonetheless, the known quantum algorithms give rise to fundamental existential threats to these cryptographic foundations. For example, while offering an exponential speedup in solving both the integer factorization problem and the elliptic curve discrete logarithm problem (ECDLP)~\cite{shor1999polynomial,kishi2025simulation}, Shor's algorithm renders the known classical signatures vulnerable. Unlike classical brute-force methods, which are computationally infeasible, a quantum adversary can derive a private key from a public key in polynomial time. Given that public keys are inherently transparent on the blockchain ledger, this vulnerability permits the unrestricted forgery of digital signatures, user impersonation, and arbitrary asset transfer. This exposure necessitates a fundamental transition from computational security assumptions to ITS standards.
Besides, the consensus layer faces systemic subversion through Grover's algorithm. This algorithm yields a quadratic speedup for unstructured search problems, reducing the complexity of finding a hash pre-image from $\mathrm{O}(2^n)$ to $\mathrm{O}(2^{n/2})$~\cite{jain2024quantum}. While increasing hash length may mitigate specific collision attacks regarding data integrity, the threat to the PoW consensus remains structural.  In a hybrid scenario where a single adversary possesses quantum capabilities while others rely on classical algorithms, this adversary could gain a disproportionate hashrate advantage. Consequently, this work advocates for a paradigm shift toward a BFT consensus mechanism, authenticated via information-theoretically secure keys, as proposed in our architecture. To provide a clear overview of these vulnerabilities and our corresponding countermeasures, Table~\ref{tab:security_comparison} summarizes the comparison of blockchain security mechanisms against quantum threats.

\begin{table*}[htbp]
    \centering
    \caption{Comparison of Blockchain Security Mechanisms against Quantum Threats.}
    \label{tab:security_comparison}
    \renewcommand{\arraystretch}{1.5}
    \newcolumntype{L}{>{\raggedright\arraybackslash}X}

    \begin{tabularx}{\textwidth}{
        >{\bfseries}l 
        L             
        L            
        L            
        L           
    }
        \toprule
        Layer / Component & 
        Classical Mechanism & 
        Quantum Threat & 
        Consequence & 
        Proposed Solution \\
        \midrule
        
        Identity \& Integrity & 
        Digital Signatures \newline (ECDSA) & 
        \textbf{Shor's Algorithm} \newline (Integer Factorization) & 
        Private key derivation; \newline Forgery \& Impersonation & 
        \textbf{ITS Authentication} \newline (via TF-QKD) \\
        \midrule
        
        Consensus Mechanism & 
        Proof-of-Work \newline (PoW) & 
        \textbf{Grover's Algorithm} \newline (Quadratic Speedup) & 
        Hashrate centralization; \newline 51\% Attack & 
        \textbf{Quantum-Secured BFT} \newline (Voting-based) \\
        \midrule
        
        Security Basis & 
        Computational Hardness Assumptions & 
        Mathematical Vulnerability & 
        Systemic Collapse & 
        \textbf{Physical Laws} \newline (Uncertainty Principle) \\
        
        \bottomrule
    \end{tabularx}
\end{table*}

\subsection{Quantum Resistance Approach}

In response to the multifaceted quantum threats challenging classical cryptographic assumptions, quantum key distribution emerges as a transformative physical-layer approach. Its guarantees are rooted in the fundamental laws of quantum physics specifically the no-cloning theorem and the uncertainty principle rather than computational hardness. This enables a comprehensive security architecture that remains secure regardless of the adversary's future computational power.

Within a quantum-secured blockchain architecture, QKD is explicitly utilized to continuously distribute these symmetric ITS keys among decentralized network nodes. Rather than relying on computationally vulnerable public-key infrastructures, these quantum-derived symmetric keys are employed to generate unconditionally secure message authentication codes (MAC) for transaction broadcasting and consensus voting. This mechanism provides an exact ITS alternative to traditional asymmetric digital signatures. While QKD necessitates an authenticated classical channel for initialization which is typically bootstrapped via a small pre-shared key, it effectively shifts the system's authentication foundation from algorithmic hardness to physical state disturbance. Consequently, integrating this QKD-enabled physical infrastructure with quantum-safe voting protocols, such as BFT, establishes a mathematically and physically robust framework that rigorously secures both the transaction and consensus layers.

Pioneering works~\cite{kiktenko2018quantum,yang2024qbma,reddy2025quantum} have experimentally demonstrated the QKD-secured blockchains over urban fiber networks. However, these seminal systems typically rely on point-to-point BB84 protocol. A limitation of such architectures is the requirement for a direct physical quantum link between every pair of communicating nodes. In a distributed network, it requires a fully connected mesh topology, where the number of required optical fiber links scales quadratically ($\mathrm{O}(N^2)$) with the number of nodes. This hardware complexity imposes severe constraints on network scalability and cost-efficiency.

To address current scalability limitations, we propose a TF-QKD-based framework to extend the geographical reach of blockchain networks. This approach aims to contribute to quantum network design by mitigating potential challenges for implementations.

Intriguingly, the framework ensures immunity to detector side-channel attacks through a MDI architecture. Unlike the traditional BB84 protocol, where security is susceptible to imperfections in receiving detectors, our approach involves terminal nodes (such as Alice and Bob) transmitting optical pulses to an untrusted central relay (Charlie) for interference measurement. Consequently, security relies exclusively on the correlations between the prepared states and the relay's public outcomes, maintaining system integrity even if the relay's detectors are imperfect or controlled by an adversary.

From an MDI-driven structural perspective, TF-QKD facilitates a star-shaped physical topology, where multiple blockchain nodes connect to a single relay via optical links. This configuration reduces physical complexity from a quadratic $\mathrm{O}(N^2)$ dependence to a linear $\mathrm{O}(N)$ scale, thereby minimizing deployment costs and simplifying network expansion. Moreover, while the physical quantum layer utilizes a centralized relay, the upper-layer blockchain consensus retains its logical decentralization. Of note, the secret key rate scales with the square root of the channel transmittance ($\sqrt{\eta}$) rather than the linear relationship ($\eta$) observed in conventional QKD schemes.
While eliminating detector vulnerabilities, optimizing physical infrastructure, and enabling long-haul connectivity, the TF-QKD-secured blockchains provide a superior physical foundation required for large-scale quantum-secured distributed ledgers.

\section{Blockchain with Quantum-classical hybrid Architecture}
\label{sec:system_architecture}

We propose a quantum-classical hybrid architecture that decouples physical key generation from logical consensus. By adopting a TF-QKD star topology, quantum measurements are centralized at a relay. Crucially, this relay is deliberately assumed to be untrusted. Thanks to the MDI property, the relay only measures single-photon interference patterns without accessing the encoded bits, meaning it cannot extract secret keys even if fully compromised. This zero-trust physical design perfectly aligns with the blockchain's decentralized ethos by eliminating any single point of trust. Consequently, this strategy translates TF-QKD's scalability into a practical network model, providing a continuous supply of ITS keys without the hardware bottlenecks of fully connected meshes.

\begin{figure}[t] 
    \centering
    \includegraphics[width=\columnwidth]{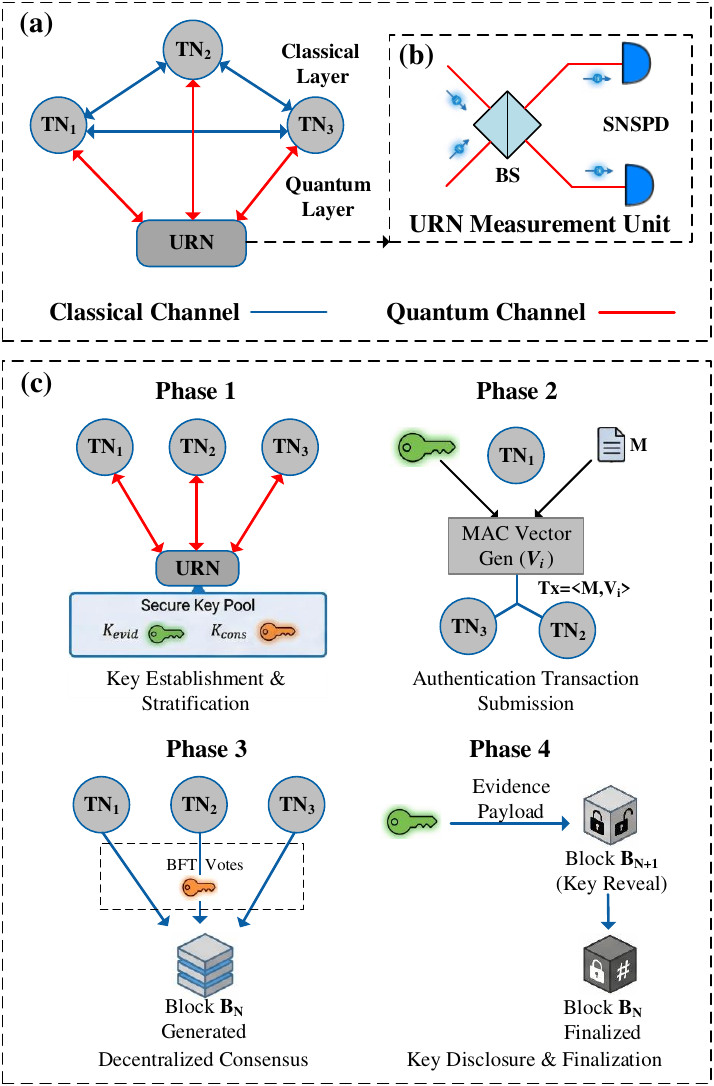} 
    \caption{{Schematic overview of the proposed quantum-resistant blockchain architecture.} \textbf{(a)} The hybrid network topology decouples the physical quantum layer from the logical classical layer. There are three terminal nodes connected to a central URN for the TF-QKD interference, while consensus communications occur over a peer-to-peer classical mesh. \textbf{(b)} The measurement of URN. BS: beam splitter; SNSPD: superconducting nanowire single photon detector. \textbf{(c)} The four-phase operational workflow: \textbf{Phase 1} performs continuous key stratification into Evidence Keys ($K_{\text{evid}}$) and Consensus Keys ($K_{\text{cons}}$); \textbf{Phase 2} employs $K_{\text{evid}}$ for Wegman-Carter authenticated transaction submission; \textbf{Phase 3} executes decentralized BFT consensus using $K_{\text{cons}}$; \textbf{Phase 4} achieves block finalization and reveals $K_{\text{evid}}$ to ensure non-repudiation.}
    \label{fig:system_architecture}
\end{figure}

\subsection{Scheme Design of Quantum Blockchain}

The quantum-secured blockchain can be designed with a distributed star architecture, which involves the TF-QKD-enabled physical layer and the blockchain-embedded logical consensus layer, as shown in Figure \ref{fig:system_architecture}(a). It can be described respectively as follows.

\textbf{1) TF-QKD-enabled Physical Layer:} 

For the physical layer, a centralized star-topology network is elegantly employed for the TF-QKD protocol. Structurally, the infrastructure consists of spatially distributed terminal nodes (TN) connected via dedicated quantum channels links to a centralized untrusted relay node (URN). Without loss of generality, we consider the terminal nodes $\{\rm{TN}_i: i\in \{1,2,\cdots N\}\}$, in the distributed blockchain. To establish secure cryptographic keys, each node $\rm{TN}_i$ is equipped with a stabilized laser source, intensity modulators for decoy state generation, and phase modulators for encoding cryptographic bits and bases. To ensure the required phase coherence across spatially separated terminal nodes, a continuous-wave reference laser is multiplexed with the quantum signals to provide a global phase reference, enabling active phase-tracking and compensation for fiber-induced phase drifts. Furthermore, the intensity modulators generate distinct intensity levels (decoy states) to rigorously bound the yield of single-photon emissions and mitigate the potential photon-number-splitting (PNS) attacks.

During operation, each $\rm{TN}_i$ independently prepares and transmits these phase-locked weak coherent pulses to URN, where it functions strictly as an analog interferometer, primarily comprising a 50:50 optical beam splitter and highly efficient single-photon detectors, as shown in Figure~\ref{fig:system_architecture}(b). When the optical pulses from participants $\rm{TN}_i$ and $\rm{TN}_j$, $\forall i,j\in\{1,2, \cdots N\}$, arrive at the beam splitter, they undergo single-photon interference. Specifically, the relative phase difference between the incoming pulses dictates the detection probabilities at URN's outputs. Constructive interference triggers a click in one detector, mapping to a correlated bit configuration (e.g., a relative phase of $0$), while destructive interference triggers the other, mapping to an anti-correlated configuration (e.g., a relative phase of $\pi$). A detection event reveals only this relative phase difference without disclosing the absolute phase information encoded by either individual terminal node. 
Subsequently, URN publicly broadcasts these interference outcomes. Based on announcements, $\rm{TN}_i$ and $\rm{TN}_j$ engage in authenticated classical post-processing to distill pairwise, information-theoretically secure secret keys. This data post-processing begins with a data sifting step, where both nodes announce their encoded bases over an authenticated classical channel, discarding instances with mismatched bases or invalid URN measurements. Following this, rigorous parameter estimation is performed using the decoy states to quantify the quantum bit error rate (QBER) and phase error rate. Finally, they execute error reconciliation and privacy amplification. Because the central relay merely records interference patterns, the physical layer achieves the known MDI security~\cite{mdiqkd}, fundamentally eliminating all detector-side channel vulnerabilities.

\textbf{2) Blockchain-QKD Hybrid Logical Consensus Layer:} 

The logical consensus layer implements a decentralized and permitted consortium blockchain protocol. This layer is designed to fuse the continuous keys generated from the QKD-based physical layer with the distributed ledger's rigorous cryptographic authentication requirements. While the physical key generation is anchored to a centralized URN, the logical blockchain consensus operates over a fully decentralized, peer-to-peer classical overlay network that directly interconnects all terminal nodes. This consensus mechanism encompasses critical operations such as transaction broadcasting, ledger synchronization, and block validation. This architectural decoupling represents a critical design choice because it ensures that the central URN does not become a single point of failure or a communication bottleneck for classical consensus liveness. The technical integration between these two distinct domains is mediated by local key management buffers at each terminal node. These buffers harvest the continuous stream of symmetric keys distilled from the TF-QKD layer, acting as a dynamic reservoir to secure communications of all subsequent classical ledgers.

To address the multi-layered security challenges posed by quantum computing, computationally vulnerable asymmetric digital signatures are entirely replaced. Accordingly, the logical consensus layer can enforce identity authentication and message integrity using information-theoretically secure Wegman-Carter message authentication codes (MAC)~\cite{degabriele2024sok}. When a terminal node $\rm{TN}_i$ initiates a transaction or broadcasts state-machine replication messages during the BFT consensus process (such as the prepare and commit phases), the plaintext payload is authenticated using a global MAC vector. This vector is constructed dynamically using strict one-time-use keys drawn from the fresh TF-QKD key pool shared pairwise between the broadcasting $\rm{TN}_i$ and all other validator nodes $\rm{TN}_j$, $j\neq i$, in the consortium. By embedding these quantum-derived authenticators directly into the peer-to-peer broadcast routing mechanism, this network design effectively translates the physical security guarantees of quantum mechanics into logical data integrity, ensuring systemic non-repudiation across the distributed network.

Notably, compared with traditional blockchains that rely on computational hardness assumptions and may become vulnerable to future quantum algorithmic advances, this hybrid architecture lies in the deliberate decoupling of the physical key generation from the logical consensus process. 
While pioneering quantum-secured blockchains have sought to address this by using QKD, they frequently entangle the quantum communication infrastructure directly with the decentralized ledger topology, inadvertently introducing the deployment bottlenecks. Whereas, the proposed scheme attempts to separate these concerns. By physically centralizing the complex quantum interference processes at URN while strictly preserving the decentralized peer-to-peer structure of the BFT consensus at the logical consensus layer, this architecture offers a pragmatic bridge. It does not seek to alter the core philosophy of distributed ledgers; rather, it cautiously aims to provide a provably secure cryptographic substrate that can be smoothly integrated with consortium blockchain operations, maintaining logical decentralization without demanding a fully decentralized quantum physical infrastructure.

\subsection{Implementation of Quantum Blockchain} 

In what follows, we implement a layered protocol stack to secure the interactions within this hybrid framework. To clarify the operational scope, the system is deployed over a consortium network consisting of $N$ terminal nodes. The consensus layer implements a quantum-secured BFT mechanism capable of tolerating up to $f$ malicious nodes ($N = 3f+1$). Within this protocol, a standard transaction request is structured as a tuple $Tx_{\text{req}} = \langle ID_i, \text{Ct}_i, M, \mathcal{V}_i \rangle$, encapsulating the sender's identity, a monotonically increasing counter (nonce), the message payload, and a global MAC vector for ITS authentication. Supported by the high-rate TF-QKD physical layer, this architecture is engineered to sustain scalable transaction throughput, demonstrating a capacity of hundreds of transactions per second (TPS) for typical metropolitan consortiums (with a rigorous cross-layer throughput evaluation detailed in Section~\ref{sec:performance}).

The detailed algorithm of the TF-QKD-based quantum blockchain can be found in Appendix~\ref{app:protocol_workflow}. As shown in Figure~\ref{fig:system_architecture}(c), the implementation of the operational workflow involves the physical key establishment process (Phase 1), the authenticated transaction submission (Phase 2), the decentralized consensus (Phase 3), and the key disclosure (Phase 4).

\textbf{Phase 1: Pairwise Key Establishment and Stratification }

The security foundation of the proposed architecture relies on the continuous generation of ITS keys, which serve as the cryptographic anchor for authenticating classical communications. This phase is dedicated to the establishment and lifecycle management of pairwise secret keys between $\rm{TN}_i$ and $\rm{TN}_j$. 

The key generation process in TF-QKD operates through a distinct separation of physical interference and data post-processing. For an arbitrary pair of nodes $\mathrm{TN}_i$ and $\mathrm{TN}_j$, the process initiates with the independent preparation of phase-encoded weak coherent pulses, which are transmitted over dedicated optical fiber channels to the central URN. The URN functions solely as an interferometer, performing the single-photon measurements to reveal relative phase correlations without accessing the specific encoded bit values. Following the public broadcast of measurement outcomes, $\mathrm{TN}_i$ and $\mathrm{TN}_j$ engage in a classical post-processing stage. We note that this stage, which comprises parameter estimation, error reconciliation, and privacy amplification, is conducted over an authenticated classical channel. This ensures that the final distilled key, $K_{ij}$, remains unknown to the URN.

To ensure robust system availability, key establishment is orchestrated as an asynchronous background process designed to maintain a dedicated secure key pool for each node pair. Furthermore, each terminal node $\mathrm{TN}_i$ implements a resource-aware scheduling strategy that actively monitors local key reserves. This mechanism prioritizes quantum transmission slots for peers whose key pools deplete below a predefined safety threshold, thereby striving to guarantee that a sufficient buffer of ITS keys is available to meet the authentication demands of the consensus layer.

We suggest a dual-key stratification strategy  which necessitates key disclosure and the imperative of blockchain immutability which mandates strict key secrecy.
The first key stream, designated as evidence keys $K_{\text{evid}}$, is employed exclusively for the authentication of transaction payloads $M$ between terminal nodes $\mathrm{TN}_i$ and $\mathrm{TN}_j$. They underpin the generation of MAC vectors and act as the foundation for the delayed key disclosure mechanism. By eventually revealing $K_{\text{evid}}$ after block finalization, the system transforms symmetric authentication into publicly verifiable evidence, thereby ensuring non-repudiation.
Whereas, the second stream comprises consensus keys $K_{\text{cons}}$. They are strictly confined to the consensus layer, serving to authenticate block headers and validate voting messages during the prepare and commit phases of the BFT protocol. To preserve the structural integrity of the ledger history, $K_{\text{cons}}$ are subject to a zero-disclosure policy. They are never revealed to the public or the untrusted relay and are securely erased immediately following the verification process. This cryptographic bifurcation ensures that while the validity of individual transactions becomes transparent and publicly auditable post-disclosure, the blockchain's consensus history remains shielded by the undisclosed keys, rendering the chain immune to long-range forgery attacks.

\textbf{Phase 2: Authenticated Transaction Submission via MAC Vectors}

To guarantee source authenticity and non-repudiation within the broadcast network, we employ a vector of authenticators scheme underpinned by the evidence keys, $K_{\text{evid}}$. It is designed to enforce the strict cryptographic synchronization across the distributed nodes while providing ITS for transaction requests. Given that the WC authentication scheme necessitates the strict one-time usage of secret keys~\cite{degabriele2024sok}, the precise state synchronization between the sender and the verifiers is a prerequisite for achieving stability of the system. 

For a network consisting of $N$ nodes, each terminal node $\mathrm{TN}_i$, $\forall i\in\{1,2,\cdots, N\}$, maintains a monotonically increasing the local transaction counter, $\text{Ct}_i$. It acts as a unique nonce to prevent replay attacks, and simultaneously functions as a deterministic pointer to the index of the generated key within the pre-shared key pool.
When $\mathrm{TN}_i$ initiates a transaction with a payload $M$ (comprising transfer details and timestamps), it constructs a global authentication vector $\mathcal{V}_i$ to cryptographically bind the message to its origin and sequence. After that, $\mathrm{TN}_i$ retrieves the specific key pair $(k_{\text{hash}}, k_{\text{otp}})$ corresponding to $\text{Ct}_i$ from the secure pool shared with each peer $\mathrm{TN}_j$, $\forall j \in \{1, \dots, N\}$. The authentication tag $\tau_{i,j}$ is then computed by concatenating the counter with the message payload. This concatenation ensures that the authentication tag is inextricably linked to the specific transaction instance, thereby preventing packet reordering attacks. The mathematical formulation of the tag for each peer $j$ is expressed as:
\begin{equation} \label{eq:mac_computation}
    \tau_{i,j} = 
    \begin{cases} 
    h_{k_{\text{hash}}}(M || \text{Ct}_i) \oplus k_{\text{otp}} & \text{if } i \neq j \\
    \mathbf{0} & \text{if } i = j 
    \end{cases}
\end{equation}
where $h_{k_{\text{hash}}}$ represents the universal hash function selected from the $\epsilon$-ASU family, and $\mathbf{0}$ serves as a null placeholder for the sender's own position in the vector. The resulting authentication vector is denoted by $\mathcal{V}_i = [\tau_{i,1}, \tau_{i,2}, \dots, \tau_{i,N}]$.

Following the vector construction, the transaction data is encapsulated into a formal protocol data unit, denoted as $Tx_{\text{req}} = \langle ID_i, \text{Ct}_i, M, \mathcal{V}_i \rangle$. In contrast to the centralized relay architectures, this packet is broadcast via the classical authenticated channel directly to the peer network, targeting the current consensus leader (a designated terminal node in the BFT protocol) rather than the central relay. While the URN facilitates the physical generation of quantum keys in the preceding phase, it is logically excluded from this transaction submission process. The validation of the transaction format and the aggregation of $Tx_{\text{req}}$ into a candidate block are exclusively performed by the consensus leader. By decoupling the classical transaction flow from the quantum relay, the architecture mitigates the risk of censorship or attacks of URN, ensuring that the integrity of the data relies solely on the consensus of the terminal nodes.

\textbf{Phase 3: Decentralized Consensus and Verification}

Upon the dissemination of the candidate block $B_{\text{cand}}$ by the designated consensus leader (a role rotating among the terminal nodes), the network engages in a decentralized verification protocol. To achieve robust agreement amidst potential malicious behavior, the system adapts the BFT mechanism over the authenticated classical channel. The network model postulates $n$ terminal nodes accommodating at most $f$ Byzantine adversaries, satisfying the constraint $n = 3f+1$. The primary objective of this phase is to ensure that all honest nodes converge on an identical decision regarding the validity of the block proposed by the current leader, independent of the physical state of the quantum relay.

The validation process executed by each receiving node $\mathrm{TN}_j$ is rigorous and twofold, enforcing both cryptographic integrity and transactional semantics. First of all, the node verifies the authentication tags. Using the index $\text{idx}_{sj}$ extracted from the received vector, $\mathrm{TN}_j$ retrieves the corresponding one-time key pair from its local secure pool shared with the purported sender $\mathrm{TN}_s$. The authentication tag is then recomputed locally for the message concatenation $(\text{idx}_{sj} \,||\, M_k)$ following the WC construction in Equation~(\ref{eq:mac_computation}). A divergence between the computed and received tags, or the absence of a required authenticator, necessitates the immediate rejection of the specific transaction. Subsequently, contingent upon successful authentication, $\mathrm{TN}_j$ validates the semantic legitimacy of the transaction content $M_k$ against its local copy of the ledger state, screening for violations such as double-spending or insufficient balances. A candidate block $B_{\text{cand}}$ is deemed valid and vote-worthy only if every constituent transaction successfully traverses this dual-verification sequence.

Following a successful validation, $\mathrm{TN}_j$ broadcasts its endorsement by transmitting an authenticated $Accept$ vote for $B_{\text{cand}}$ directly to its peers via the classical overlay network. Finality is achieved through a quorum-based mechanism; a node considers the block irrevocably finalized once it has collected and verified a set of at least $2f+1$ distinct and valid votes. This threshold mathematically guarantees that if any honest node commits to a block, the agreement is consistent across the honest majority, thereby preserving the safety property of the ledger.

To sustain network liveness against a faulty or malicious consensus leader such as one that ceases to propose blocks or selectively censors valid transactions, the protocol incorporates a robust timeout mechanism. Should a Terminal Node fail to verify and finalize a block within a deterministic time window, a view-change protocol is triggered. In this event, the terminal nodes utilize their secure peer-to-peer channels to establish a quorum regarding the failure of the current leader and deterministically elect a successor for the subsequent round. Crucially, since this consensus signaling occurs over the classical layer, the logical robustness of the blockchain is preserved even in scenarios where the URN encounters physical service interruptions, provided that the local key buffers at the terminal nodes remain sufficient to authenticate the view-change messages.

\textbf{ Phase 4: Block Finalization and Inter-Block Key Disclosure}

This phase is initiated once the consensus protocol confirms the agreement on the candidate block, denoted as $B_N$. Upon receiving and verifying a quorum of $2f+1$ authenticated commit messages from the peer nodes, the terminal node promotes the candidate block to a finalized state. Subsequently, the node computes the cryptographic hash of the block header, $H(B_N)$, which incorporates the hash of the preceding block to enforce the chain-like structure. The hash value serves as a unique identifier for the block state, ensuring that the ledger history is cryptographically immutable and deterministic across all honest nodes.

While the consensus mechanism guarantees ledger consistency, the exclusive reliance on symmetric QKD keys for authentication introduces a challenge regarding public non-repudiation. To resolve this, the architecture incorporates an inter-block key disclosure mechanism, operating on a $Commit$-$then$-$Reveal$ principle~\cite{848446}. Following the finalization and local storage of block $B_N$, the protocol mandates that every sender $\mathrm{TN}_i$ who successfully executed transactions within this block must disclose the specific OTP keys (part of the key pairs of QKD) used to generate the authentication vectors. To optimize network bandwidth, $\mathrm{TN}_i$ aggregates these keys into a compact evidence payload. In contrast to the relay-dependent approaches, this payload is broadcast directly via the classical consensus channel to the network. It is then prioritized for inclusion in the body of the subsequent block, $B_{N+1}$, by the elected leader of the next consensus round.

Meanwhile, the temporal separation between key utilization and key disclosure enhances security of the system. Since the keys for block $B_N$ are propagated only after the block's content has been hash-locked and immutably recorded by the honest majority, neither a malicious leader nor an external adversary can retroactively utilize the disclosed keys to forge or modify the transactions in $B_N$. The inclusion of the evidence payload in $B_{N+1}$ transforms the private, symmetric verification of the previous block into a publicly auditable record. Consequently, any network participant or external auditor can extract the keys from $B_{N+1}$ to re-verify the MAC vectors stored in $B_N$. As the valid construction of a global MAC vector requires the simultaneous possession of pairwise keys shared with all nodes, a condition only the legitimate sender can satisfy at the time of transmission—the successful verification of these vectors against the revealed keys provides robust cryptographic evidence of the sender's identity.

Of note, a challenge of this hybrid architecture is to establish the relation of the redesigned blockchain and the TF-QKD-based distributed network. To achieve quantum resistance, the classical blockchain is suitably modified to replace asymmetric digital signatures with MAC vectors and a delayed key disclosure mechanism. In addition, the TF-QKD structure effectively delegates detection processes to URN, thereby rendering the system inherently immune to detector side-channel attacks. Besides, its ability to overcome the linear rate-loss limit allows for scalable key distribution suitable for the star-shaped topology of the proposed network. Additionally, the ability of TF-QKD to overcome the PLOB bound $\mathrm{O}(\sqrt{\eta})$ generates the abundant cryptographic entropy required to sustain the high consumption of the MAC vectors, which will be shown in Figure \ref{fig:supply_demand}. Consequently, the modified consensus protocol is uniquely suited to harness the structural benefits of TF-QKD, translating a theoretically secure mechanism into a physically scalable and practically deployable distributed ledger.

\section{Security Analysis}
\label{sec:security_analysis}

In this section, we evaluate the security resilience of the hybrid architecture under a comprehensive adversarial framework. The security analysis focuses on the composability of information-theoretic authentication and Byzantine Fault Tolerance consensus, demonstrating how the system mitigates threats spanning from quantum-computational attacks to internal protocol deviations.

\subsection{Threatened Attack Model}
\label{subsec:threat_model}

We assume an attack threatened model that encompasses both external eavesdroppers and internal adversarial participants, operating under the assumption that the adversary possesses unbounded computational power. This model can be described as follows.

\textbf{1) External Quantum Adversary}

We postulate the existence of an external adversary, Eve, equipped with unlimited computational resources, including a universal quantum computer capable of executing Shor’s and Grover’s algorithms efficiently. Under this assumption, all computational hardness primitives such as integer factorization and discrete logarithms are considered compromised. Eve is assumed to have full control over the public classical channels, granting her the capability to eavesdrop, intercept, and replay classical messages. However, her interaction with the quantum channel is strictly constrained by the fundamental laws of quantum mechanics. Specifically, the no-cloning theorem guarantees that any attempt by Eve to measure or copy the quantum states transmitted between the terminal nodes and URN introduces statistically detectable disturbances, thereby preventing silent eavesdropping on the key generation process.

\textbf{2) Internal Adversarial Participants}

Regarding the participants of network, we distinguish between the roles of the central relay and the decentralized terminal nodes, assigning distinct adversarial capabilities to each.
On the one hand, we model the URN as a malicious entity responsible solely for the quantum physical layer. The URN may attempt to deviate from the TF-QKD protocol by falsifying single-photon measurement results or conducting Denial-of-Service (DoS) attacks on the quantum channel to inhibit key generation. However, crucially, due to the decoupled dual-layer network architecture, URN is assumed to have no control over the classical peer-to-peer consensus layer. Consequently, while the URN can attempt to exhaust the system's key reservoir, it cannot suppress, modify, or censor the classical consensus messages exchanged directly between Terminal Nodes.
But on the other, we consider the terminal nodes, which execute the BFT consensus protocol. We adopt the standard Byzantine failure model, assuming that an adversary may corrupt up to $f$ nodes in a network of $n$ participants, satisfying the constraint $n = 3f+1$. These Byzantine nodes may collude, behave arbitrarily, or go offline. Explicitly, this threat model includes the scenario of a malicious Consensus Leader. A compromised leader may attempt to undermine the ledger's integrity or liveness by proposing invalid blocks, censoring specific transactions, or equivocating (broadcasting conflicting blocks to different peers). The security of the system relies on the honest majority ($n-f$) to detect such malfeasance and trigger the view-change protocol to replace the faulty leader, ensuring the system's resilience against authority-based attacks.

\subsection{Immunity to Cryptographic Attacks}
\label{subsec:quantum_attacks}

The security of the hybrid architecture is due to decoupling the system's integrity from computational hardness assumptions, thereby neutralizing threats posed by known quantum algorithms. This subsection delineates the mechanisms by which the architecture addresses the specific vulnerabilities introduced by Shor's and Grover's algorithms.

The primary threat to transaction authenticity stems from Shor's algorithm, which can solve the ECDLP in polynomial time, rendering insecurity of the conventional digital signatures (e.g., ECDSA)~\cite{shor1999polynomial}. To circumvent this vulnerability, we substitute computational signatures with the WC-based MAC scheme. The security of this scheme can be derived from information-theoretic principles rather than computational complexity. For a message $M$, the authentication tag is computed by combining a universal hash function with a OTP:
\begin{equation}
    T_{\text{ag}} = H_{k_1}(M) \oplus k_2,
\end{equation}
where $H_{k_1}$ is selected from an $\epsilon$-Almost Strongly Universal ($\epsilon$-ASU) hash family using a secret key $k_1$, and $k_2$ is a one-time encryption key. The $\epsilon$-ASU property guarantees that the collision probability for any distinct message pair is bounded by a negligible $\epsilon$~\cite{krawczyk1994lfsr}. Crucially, the subsequent XOR operation with $k_2$ provides perfect secrecy for the hash output. Even assuming an adversary with unbounded computational power intercepts the pair $(M, T_{\text{ag}})$, the OTP ensure that no information regarding $k_1$ or the tag of any subsequent message is leaked. Consequently, the probability of a successful forgery is strictly bounded by the combinatorial properties of the hash family (exponentially small relative to the tag length) rather than the adversary's computing capabilities. A formal derivation of this bound is detailed in Appendix \ref{app:wc_security}.

It is imperative to note that the unconditional security of this construction is predicated on the strict $one$-$time$ usage of the key $k_2$. As key reuse would expose the system to linear cryptanalysis, an adversary could derive the hash output via XOR linearity (e.g., $T_{\text{ag}_1} \oplus T_{\text{ag}_2} = H(M_1) \oplus H(M_2)$), potentially leading to compromise of system. This constraint underscores the critical necessity of the underlying TF-QKD layer. Unlike classical systems where key distribution often forms a bottleneck, the high-rate TF-QKD ensures a continuous supply of provably secure entropy. This capability allows the system to strictly enforce a unique key policy for every transaction, thereby sustaining the theoretical security guarantees in a practical deployment.

In parallel to the cryptographic integrity, the system also addresses threats to the consensus mechanism. Classical blockchains utilizing Proof-of-Work are vulnerable to Grover's algorithm, which affords a quadratic speedup in unstructured search problems, effectively lowering the hashrate barrier for a 51\% attack. The proposed architecture is intrinsically immune to this vector by adopting a voting-based BFT consensus rather than a puzzle-based mechanism. Since finality is achieved through authenticated communication quorums rather than brute-force pre-image search, there is no computational search space for a quantum adversary to accelerate. The integrity of the ledger, therefore, relies solely on the honest majority assumption and the authentication of the voting messages, remaining robust regardless of the adversary's quantum computational advantage.

\subsection{Resilience to Internal Adversaries}
\label{subsec:internal_robustness}

The system is explicitly engineered to maintain safety, liveness, and auditability despite the presence of malicious internal components. Accordingly, this analysis focuses on the distinct constraints imposed on the physical relay, the Byzantine tolerance of the consensus layer, and the cryptographic enforcement of non-repudiation.

Regarding the role of URN, the proposed dual-layer architecture strictly limits its adversarial capabilities， as URN is restricted solely to the generation of raw keys. Critically, it is physically excluded from the classical peer-to-peer network where transaction broadcasting and consensus voting occur，and thus possesses no knowledge of the distilled pairwise secret keys ($K_{ij}$). Consequently, URN is incapable of modifying transaction payloads, forging consensus votes, or censoring classical traffic. While the malicious URN attempts a DoS attack by halting the quantum key distribution, this only impacts the replenishment of the key reservoir. The consensus protocol remains operational using buffered keys, and the system retains the capacity to trigger failover mechanisms without  URN's participation.

Meanwhile, the resilience of system extends to the corruption of consensus participants, including the scenario of a malicious consensus leader. We adopt the standard BFT assumption where the number of faulty nodes $f$ satisfies the constraint $n = 3f+1$. A compromised leader may attempt to compromise the ledger's integrity by broadcasting invalid blocks or attempting double-spending attacks. However, the validity of any block is predicated on the accumulation of a quorum of $2f+1$ cryptographically authenticated votes. Since the leader cannot forge the pairwise MAC signatures of honest nodes, any attempt to fabricate a quorum or alter the immutable history will be detected by the honest majority. Furthermore, should the leader exhibit liveness failures (such as censoring valid transactions or refusing to propose blocks), the view-change protocol, executed over the authenticated peer-to-peer network, ensures that the leadership is transferred to a correct node, thereby guaranteeing liveness of continuous protocol.

A critical challenge in symmetric-key systems is to achieve non-repudiation without digital signatures. Our architecture addresses this through the mechanism of information asymmetry. Consider a scenario where a malicious receiver (e.g., Bob) attempts to frame a sender (e.g., Alice) by forging a broadcast message. In our vector-based authentication scheme, a valid broadcast requires a vector containing correct authentication tags for every node in the network. However, Bob possesses knowledge only of the specific key $K_{AB}$ shared with Alice; he has zero knowledge of the keys Alice shares with other peers (e.g., $K_{AC}$ and $K_{AD}$). Therefore, to successfully forge a global broadcast that passes verification by the network majority, Bob would need to simultaneously guess the secret keys of all other nodes, a feat that is statistically impossible absent a total collusion of the network. This asymmetry ensures that a valid MAC vector serves as undeniable proof of the sender's origin.

Strikingly, to defend against the long-range attacks where an adversary utilizes historically disclosed keys to forge an alternative blockchain history~\cite{deirmentzoglou2019survey}, we implement a strict dual-key stratification strategy. Although the evidence keys ($K_{\text{evid}}$) are eventually disclosed to facilitate public auditing, the structural integrity of the blockchain is anchored by the consensus keys ($K_{\text{cons}}$). These keys, used exclusively for authenticating block headers and linking the cryptographic chain, are never disclosed and are securely erased immediately after block finalization. This $forward$-$secrecy$ property ensures that even if an adversary gains full access to all historical transaction keys, they remain mathematically incapable of constructing valid block headers to extend a forged chain, thereby preserving the immutability of the ledger.

\begin{figure}[t]
    \centering
    \includegraphics[width=\linewidth]{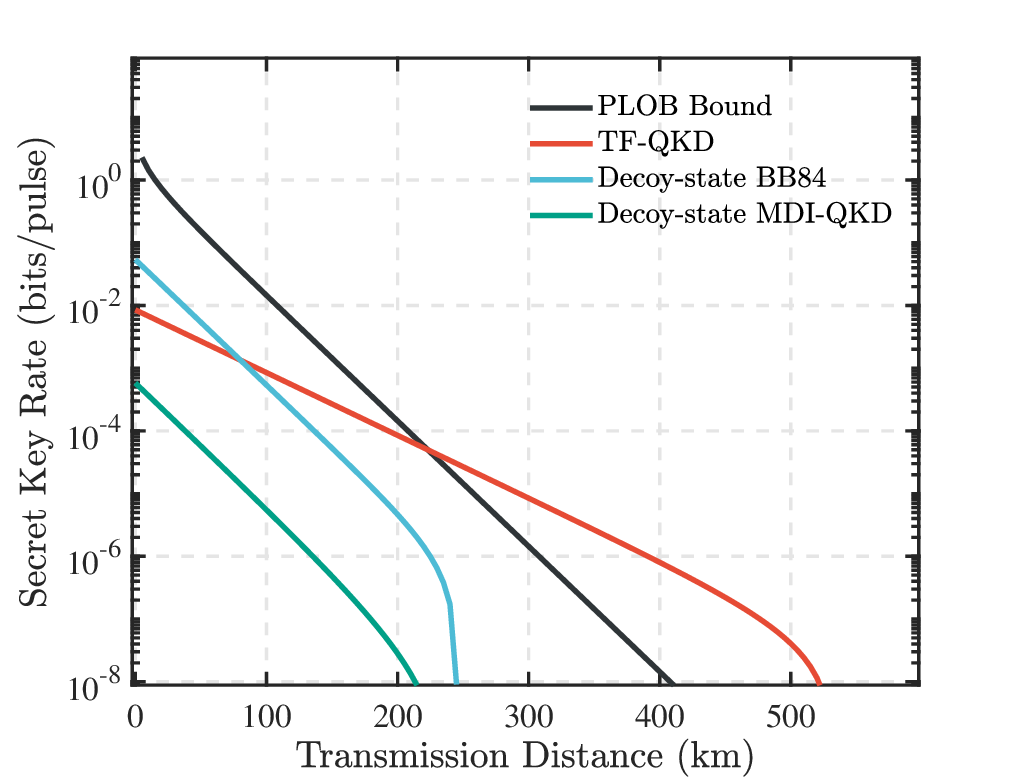}
    \caption{{Performance comparison of secret key rates.} The TF-QKD protocol (red solid line) overcomes the PLOB bound (black solid line) and the distance limitations of MDI-QKD (green solid line). While BB84 (blue solid line) shows higher rates at short distances, it lacks the measurement-device-independent security feature inherent to the TF-QKD architecture.}
    \label{fig:performance_comparison}
\end{figure}

\subsection{Practical Security of Physical QKD Layer}
\label{subsec:qkd_security}

The foundational security of the proposed architecture is anchored in the information-theoretic guarantees provided by the TF-QKD protocol. Unlike the classical cryptographic primitives, the security of this physical layer is derived from the fundamental postulates of quantum mechanics rather than computational complexity assumptions. In the implemented scheme, legitimate terminal nodes prepare and transmit phase-encoded weak coherent pulses to the central URN. The protocol security relies on the physical reality that any eavesdropping attempt on the quantum channel inevitably disturbs the single-photon interference pattern observed at the untrusted relay. By rigorously monitoring channel parameters (i.e., transmittance and phase error rates), the communicating parties can bound the potential information leakage to an adversary. Through classical data post-processing, including parameter estimation and privacy amplification, a final secret key $K_{ij}$ is distilled. This process ensures that the adversary's knowledge of the final key is exponentially suppressed, providing a provably secure basis for the subsequent authentication layer.

Beyond the theoretical security, the practical robustness of the physical implementation is of equal critical importance. A comparative analysis of security boundaries is illustrated in Figure~\ref{fig:performance_comparison}. While the traditional decoy-state BB84 protocol~\cite{lo2005decoy} exhibits superior key generation rates in short-to-medium haul scenarios (e.g., distances less than 200 km), its security model is predicated on the assumption of trusted measurement devices. This dependency leaves the system exposed to sophisticated detector side-channel attacks, such as detector blinding or efficiency mismatch exploitation, which target imperfections in physical detection hardware.

Intriguingly, the adoption of TF-QKD in our system is a strategic decision to mitigate these specific physical vulnerabilities. By delegating the measurement tasks to URN, the security proof treats the detection facility as a $black$ $box$, thereby removing the requirement for trusted detectors. Consequently, although the TF-QKD protocol may yield lower key rates compared to the BB84 protocol at short distances, it effectively eliminates the vector for detector-side channel attacks. Furthermore, regarding scalability, TF-QKD demonstrates an advantage in long-distance transmission. As evidenced, the TF-QKD protocol scales with the square root of channel transmittance $\mathrm{O}(\sqrt{\eta})$, surpassing both the BB84 protocol and the standard MDI-QKD protocol. This characteristic allows the architecture to overcome the PLOB bound, ensuring sufficient key volume for the OTP encryption even across extended metropolitan or inter-city links.

\begin{figure*}[t]
    \centering
    \includegraphics[width=\linewidth]{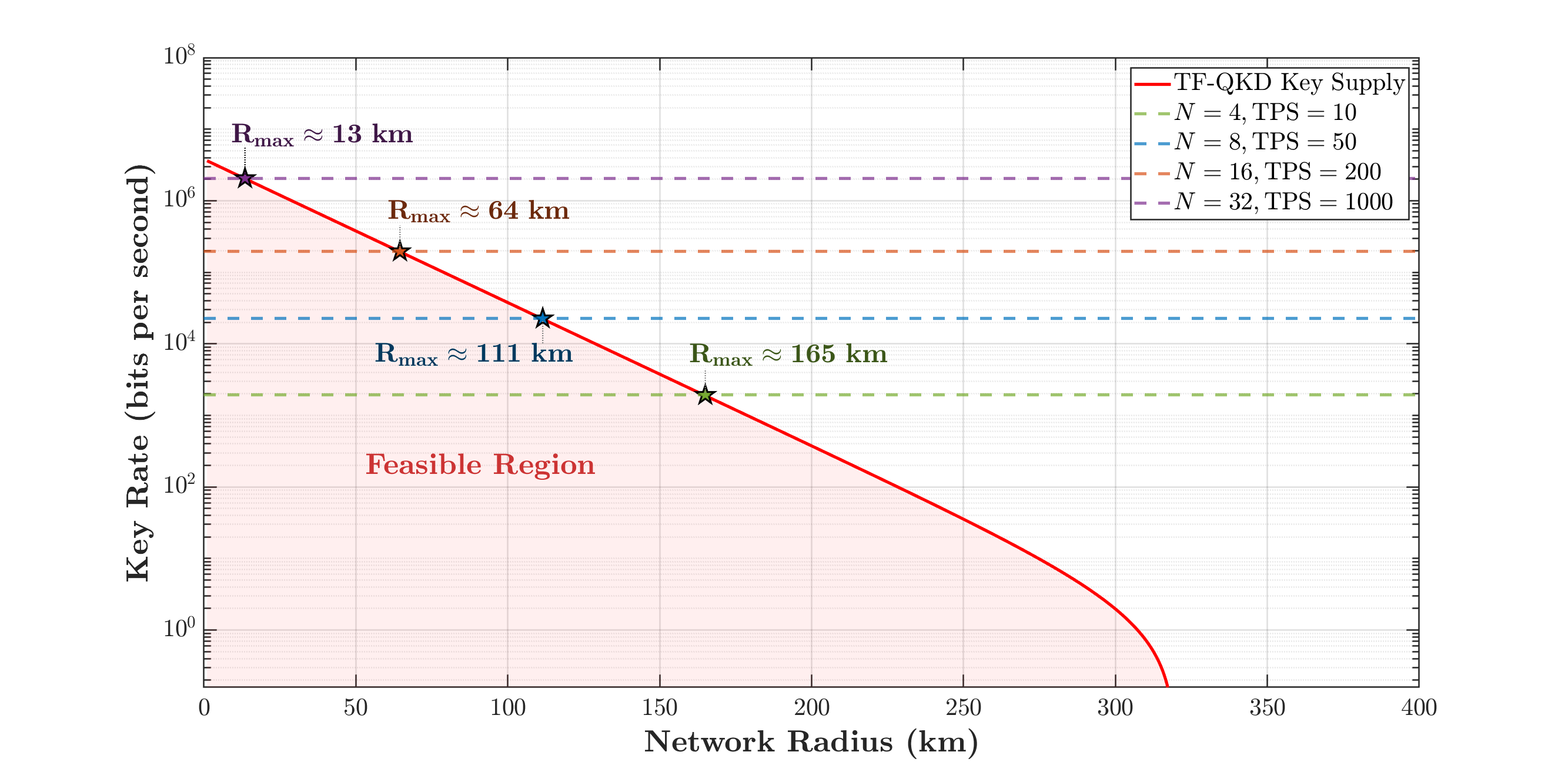}
    \caption{{Supply-demand equilibrium of the TF-QKD-secured network.} This figure demonstrates the equilibrium between the physical QKD key generation capacity (supply) and the cryptographic key consumption (demand) under various network scales ($N$). Here, the network radius refers exclusively to the QKD physical fiber distance from the decentralized terminal nodes to the central quantum relay, which inherently dictates the physical layer's key supply limits due to optical attenuation. The targeted transaction throughput (TPS) represents the cryptographically supportable capacity—the maximum number of transactions per second the network can authenticate before exhausting the continuously generated QKD keys. The intersection points $\mathrm{R}_{\max}$ mark the maximum physical deployment radius capable of sustaining the required key demand for a given consensus TPS target.}
    
    \label{fig:supply_demand}
\end{figure*}

\section{Performance Analysis}
\label{sec:performance}

In this section, we establish a comprehensive analytical framework to evaluate the implementation feasibility of the quantum-secured blockchain architecture. The analysis confirms the relation between key generation capabilities of the physical layer (supply) and cryptographic consumption requirements of the application layer (demand).

\subsection{Analytical Modeling Framework}
\label{subsec:modeling}

To evaluate capacity of the secure key generation, we adopt a system level perspective. While the fundamental link physics follows the seminal TF-QKD protocol, we denote $R_{\text{sup}}$ as the supply capacity of the aggregate secret key generation of the entire network, rather than a single peer-to-peer link rate. 
In the light of scalable network architectures with an adaptable star-topology, this definition highlights the usage of dynamic optical switching and multi-user measurement units (MU). This approach allows for parallel key generation among multiple user pairs, effectively bypassing the bandwidth bottlenecks associated with traditional time-division multiplexing.
We make an assumption that the physical layer is capable of supporting the concurrent quantum transmissions, allowing the secret key rate to scale with the square root of the channel transmittance $\mathrm{O}(\sqrt{\eta})$ across  network. Additionally, to account for an impact of environmental disturbances in realistic star-topology deployments, this model incorporates the framework of phase noise analysis established in \cite{bertaina2024phase}. As a consequence, the asymptotic secure key rate can be derived as a function of the network radius $R$ and the residual phase noise variance $\sigma_\phi$. The detailed mathematical derivation, including the channel loss model, the phase-matching conditions, and the error rate estimation, has been provided in Appendix~\ref{app:key_rate_details}.

On the one hand, the total key demand, demoted by $K_{\text{dem}}$, is driven by the cryptographic operations of the upper-layer blockchain protocol. This system utilizes information-theoretically secure (ITS) message authentication codes (MAC) for both transaction submission and consensus voting. To quantify this demand, we provide the corresponding parameters, such as the total number of terminal nodes $N$, the target transaction throughput $T$ (in TPS), the average number of transactions per block $B$, the key length required for a single MAC authentication $S_{\text{key}}$, and the average number of network-wide authentication broadcast rounds $P$ required to complete the BFT consensus for one block.

On the other hand, the total key consumption is the aggregate of the bandwidth required for transaction authentication ($K_{\text{tra}}$) and consensus authentication ($K_{\text{con}}$).
For each of the $T$ transactions generated per second, while the initiator must generate $N-1$ independent MACs for verification by all peers, we have 
\begin{equation}
K_{\text{tra}} = T (N-1)  S_{\text{key}}.
\end{equation}
Meanwhile, as the consensus process, operating at a rate of $T/B$ blocks per second, requires $P$ rounds of broadcast, we achieve
\begin{equation}
K_{\text{con}} = (T/B)  P  N  (N-1) S_{\text{key}}, 
\end{equation}
where each node communicates with $N-1$ peers.
As consequence, we obtain the unified model for the total key consumption given by
\begin{equation}
\label{eq:demand_final}
K_{\text{dem}}(N, T) = T(N-1)S_{\text{key}}\left( 1 + \frac{P N}{B} \right).
\end{equation}
It reveals that consumption scales linearly with throughput $T$ but quadratically with network size $N$.

Of note, the proposed dual-key stratification may bring a minimal overhead. Since evidence keys are batched and consensus keys are consumed per-block, the additional consumption is approximately $\Delta K \approx 1/B$. For example, taking $B=2500$ for $\Delta K \approx0.04\%$, the overhead can be negligible. As a result, we utilize Eq.~(\ref{eq:demand_final}) as the baseline of demand model.

Unexpectedly, the operational feasibility of the system is governed by the supply-demand equilibrium condition described as follows
\begin{equation}
\label{eq:equilibrium}
R_{\text{sup}}(L, \sigma_\phi) \ge K_{\text{dem}}(N, T).
\end{equation}
As confirmed, the constraint in Eq.~(\ref{eq:equilibrium}) for the large $N$ relies on the aggregate capacity assumption supported by the scalable architecture. This inequality manifests the boundaries of feasibility that will be evaluated in what follows.

\subsection{Numerical Simulations}
\label{subsec:evaluation}

Based on the analytical framework, we evaluate performance boundaries of the blockchain system in terms of the supply-demand equilibrium. In numerical simulations,  the parameters are summarized in Table~\ref{tab:sim_params}. The parameters of physical layer correspond to state-of-the-art superconducting nanowire single-photon detectors (SNSPD), while the parameters of blockchain layer are optimized for metropolitan consortiums.

It is crucial to distinguish between the logical and physical constraints of this hybrid architecture. While the logical blockchain consensus operates over a globally distributed peer-to-peer overlay, the underlying TF-QKD key distribution is strictly bounded by optical fiber attenuation. Therefore, the network radius in our evaluation defines the maximum physical distance of the star-topology quantum subnet (from terminal nodes to the central URN). Similarly, the evaluated "TPS" does not merely reflect the standard consensus bottleneck (e.g., network propagation or block validation time), but rather denotes the cryptographically constrained throughput—the maximum transaction volume that the physical layer's symmetric key supply can continuously authenticate without depletion.

\begin{table}[htbp]
    \centering
    \caption{Parameters for Numerical Simulations }
    \label{tab:sim_params}
    \begin{tabular}{lcc}
        \toprule
        \textbf{Parameter} & \textbf{Symbol} & \textbf{Value} \\
        \midrule
        \multicolumn{3}{l}{\textbf{Physical Layer}} \\
        Dark Count Rate & $P_d$ & $10$ Hz \\
        Detection Efficiency & $\eta_{\text{det}}$ & $90\%$ \\
        Fiber Attenuation & $\alpha$ & $0.2$ dB/km \\
        System Repetition Rate & $f_{r}$ & $1$ GHz \\
        \midrule
        \multicolumn{3}{l}{\textbf{Blockchain Layer}} \\
        Block Size & $B$ & $2500$ tx \\
        Auth. Tag Length & $S_{\text{key}}$ & $64$ bits \\
        Consensus Rounds & $P$ & $3$ \\
        \bottomrule
    \end{tabular}
\end{table}

\begin{figure*}[t]
    \centering
    \includegraphics[width=\linewidth]{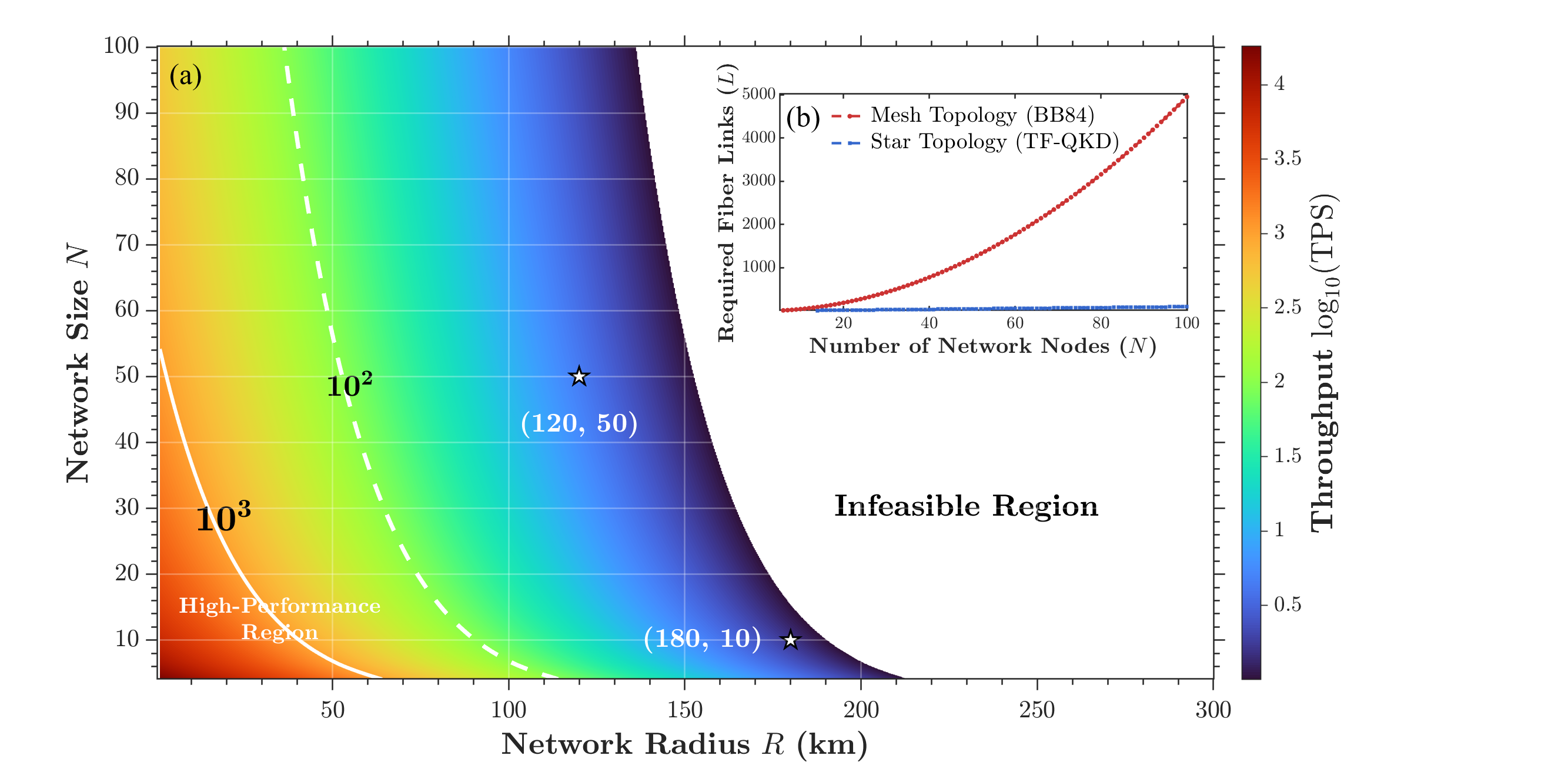}
    \caption{{Scalability and topology evaluation of the proposed architecture.} {(a) Feasibility heatmap.} Relation of network radius $R$ and node size $N$ for the variable throughput TPS ($\log_{10}$). The solid white contour represents the high-performance threshold of $10^3$ TPS, while the dashed contour marks $10^2$ TPS. Specific operational points discussed in the text are marked: the marker at $(R=120, N=50)$ illustrates the limit for metropolitan consortiums, while the marker at $(R=180, N=10)$ demonstrates feasibility for extended regional links. {(b) Quantitative comparison of physical link complexity} between the BB84-based mesh network and the TF-QKD-based star topology.}
    \label{fig:feasibility}
\end{figure*}

\begin{figure*}[t]
    \centering
    \includegraphics[width=\linewidth]{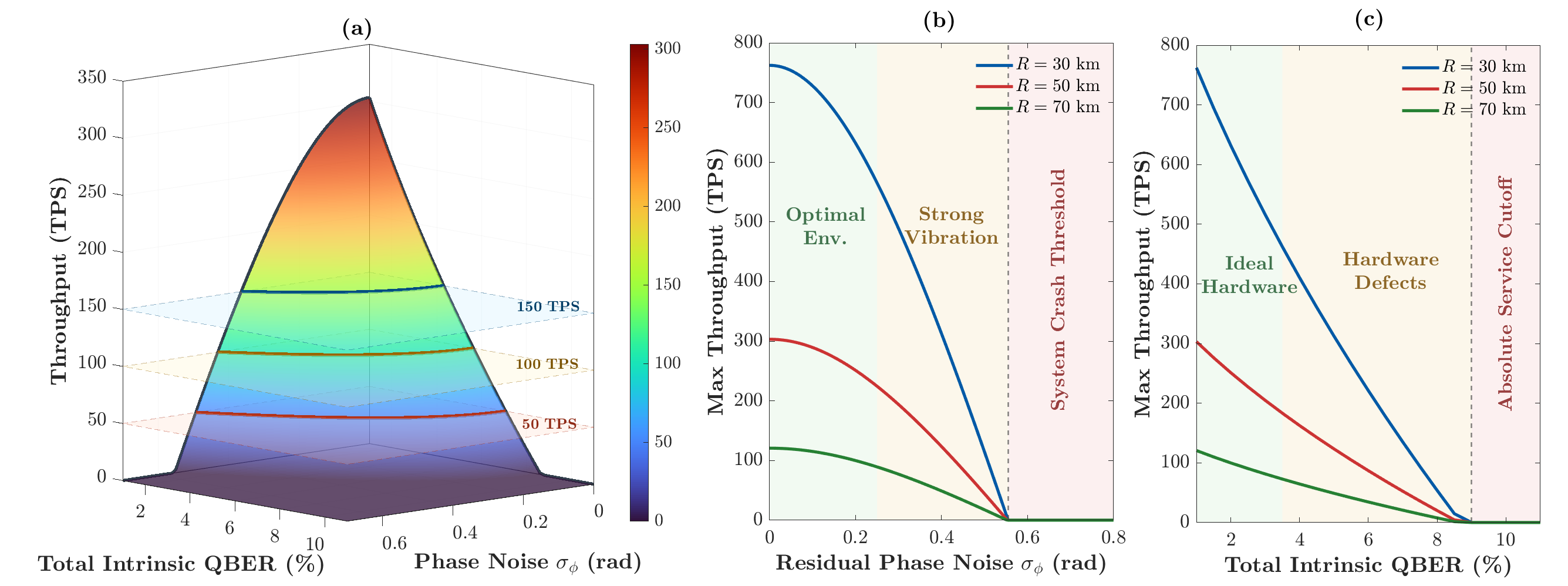} 
    \caption{{Joint performance landscape illustrating the coupled impact of phase noise and intrinsic QBER on system throughput.} 
    {(a)} The 3D surface represents the achievable TPS for a metropolitan network ($R=50$ km). 
    Horizontal semi-transparent planes are inserted at specific throughput thresholds (50, 100, and 150 TPS) to visualize service-level boundaries. 
    The intersection contours (solid lines) define the critical trade-off frontiers: to maintain a high-performance tier (e.g., more than $150$ TPS, blue plane), the system requires strict control over both environmental stability ($\sigma_\phi < 0.4$) and hardware maintenance (QBER $< 4\%$). While illustrating performance sensitivity analysis under individual physical constraints for varying network radius,
    we consider {(b)} impact of dynamic environmental phase noise ($\sigma_{\phi}$) on throughput,
    and {(c)} impact of static hardware degradation (intrinsic QBER). }
    \label{fig:single_constraint_robustness}
\end{figure*}

In Figure \ref{fig:supply_demand}, we demonstrate the relation between secret key rate of the TF-QKD system (red solid line) and  requirements of the key consumption (dashed lines) for distinct network scenarios ranging from minimal ($N=4$) to medium-scale ($N=32$) clusters, characterizing of the supply-demand equilibrium of the proposed quantum-secured blockchain system. 
The red shaded area represents the achievable key rate envelope supported by the TF-QKD system. The system is operationally feasible for a given scenario only within the range where the supply curve strictly exceeds the corresponding demand threshold. 
The intersection points, marked as $R_{\max}$, denote the maximum supportable network radius for each configuration.
The results demonstrate a perfect trade-off between capacity and coverage. 
For a lightweight, wide-area scenario ($N=4, \mathrm{TPS}=10$), the system supports an ultra-long coverage radius reaching approximately 165 km, validating its potential for inter-city backbone links. 
Intermediate configurations ($N=8$ and $N=16$) show a consistent decline in coverage to 111 km and 64 km respectively, following the scaling trend.
Conversely, for a high-throughput metropolitan scenario ($N=32, \mathrm{TPS}=1000$), the operational radius is severely constrained to approximately 14 km. The drastic reduction in operational radius from 111 km ($N=8$) to 14 km ($N=32$) is driven by the non-linear scaling conflict: while the TF-QKD key supply decays exponentially with fiber distance, the BFT consensus demand surges quadratically $\mathrm{O}(N^2)$ with the node count. Consequently, rather than a universal one-size-fits-all solution, this analysis dictates a scenario-specific deployment strategy: configuring small clusters for inter-city backbone links, while strictly restricting high-frequency, large-scale consortiums to metropolitan geographic areas.

To provide a holistic view of capabilities of the system, we synthesize the constraints of network scale and coverage radius into a sketch map to demonstrate the comprehensive feasibility of practical implementations, as shown in Figure \ref{fig:feasibility} (a). This contour plot delineates the maximum achievable TPS within the parameter space of network radius ($R$) and node count ($N$). The color gradient represents the throughput magnitude on a logarithmic scale. Two performance benchmarks are highlighted by the white contours: the solid line marks the threshold of $10^3$ TPS, defining the boundary of the explicitly labeled $High$-$Performance$ region, while the dashed line indicates the $10^2$ TPS level. The region enclosed by the axes and the solid contour represents the optimal operating zone for commercial-grade applications. Beyond the high-performance region, the results demonstrate a flexible trade-off before reaching the $Infeasible$ region (where the system fails to sustain a secure key rate). Specifically, for a medium-scale consortium chain ($N=50$), the architecture remains operational over a metropolitan area with a radius of up to 120~km, albeit with reduced throughput. Conversely, for extended regional links ($R \approx 180$~km), the system maintains feasibility for smaller clusters ($N \approx 10$), providing sufficient throughput for critical data logging or lightweight financial settlements. This analysis confirms that the star-topology TF-QKD architecture offers a scalable solution capable of meeting diverse application requirements within the identified feasible region.

While evaluating the performance boundaries of transaction throughput, it is equally crucial to quantify the physical infrastructure cost required to sustain such scalability. From the perspective of physical link complexity, the full-mesh logical key requirement inherently demands a continuous and massive supply of pairwise symmetric keys among all $N$ nodes. As illustrated in Figure~\ref{fig:feasibility}(b), traditional QKD schemes fulfilling this logical requirement would necessitate a physical mesh topology, leading to an unscalable $\mathrm{O}(N^2)$ deployment of optical fibers. Whereas, the structural fusion of our system explicitly resolves this conflict. By implementing TF-QKD, the architecture provides the required logical pairwise key distribution through a physical star topology, reducing the physical infrastructure complexity to $\mathrm{O}(N)$ links connected to a central relay. Intriguingly, the star topology reduces the link requirement by approximately 98\% for a network of 100 nodes. This 98\% reduction in link complexity is fundamentally enabled by the MDI-driven decoupling mechanism, which centralizes the sensitive detection hardware at the URN while maintaining logical peer-to-peer decentralization. Effectively, this architecture shifts the bottleneck of quantum blockchain from 'hardware-prohibitive fiber meshes' to 'scalable optical star-networks', paving the way for cost-effective urban quantum ledger deployment.

Subsequently, to characterize the combined effect of dynamic and static disturbances, we present a joint performance landscape in Figure~\ref{fig:single_constraint_robustness}(a). This visualization outlines the system's operational boundaries within the parameter space of phase noise and intrinsic QBER. The surface clarifies the pertinent trade-off of noise and QBER. We find that the tolerance of the blockchain system decreases as noise increases. This degradation is fundamentally driven by the physical limitations of the TF-QKD layer: elevated phase noise from imperfect devices amplifies the QBER. To maintain information-theoretic security, the system is forced to discard a massive fraction of raw bits during privacy amplification, which directly chokes the secure key supply. Consequently, the starved key pool can no longer sustain the high-frequency Wegman-Carter MAC vector generation required by the BFT consensus. Therefore, enforcing a strict phase noise threshold is a mandatory engineering prerequisite for field-deploying this hybrid architecture.

Based on the theoretical feasibility boundaries, we then evaluate practical performance of the system under non-ideal conditions. The analysis covers three kinds of network scales, involving a compact metro-core ring ($R=30$ km), a standard metropolitan coverage ($R=50$ km), and an extended regional link ($R=70$ km). Accordingly, the results regarding dynamic environmental disturbances and static hardware degradation are illustrated in Figure~\ref{fig:single_constraint_robustness}(b) and Figure~\ref{fig:single_constraint_robustness}(c), respectively.

Taking into account an impact of dynamic phase noise $\sigma_\phi$, which arises from environmental factors such as traffic vibrations and thermal drift, we ascertain the relation between the transaction throughput (TPS) and the residual phase noise, as shown Figure~\ref{fig:single_constraint_robustness}(b). The data suggests a correlation between network scale and noise tolerance. For the compact network ($R=30$ km, blue line), the system maintains operational capability under relatively high noise conditions ($\sigma_\phi > 0.5$ rad). As the network radius expands to 50 km (red line) and 70 km (green line), the physical link loss leads to a reduction in both peak throughput and noise tolerance. All scenarios exhibit a characteristic sharp decline once the noise variance exceeds the capacity of the phase-locking feedback loop. This provides an explanation as to why the quantum-secured blockchain can not be practically deployed at the long-transmission distance, attributed to effects of link loss on the TF-QKD system.

Complementary to dynamic noise, Figure~\ref{fig:single_constraint_robustness}(c) evaluates the system's response to static hardware degradation, modeling the cumulative effect of component aging and coupling efficiency loss (intrinsic QBER). It shows a gradual decline in performance rather than an immediate failure. The multi-scale analysis indicates that shorter links appear to offer greater tolerance to hardware imperfections. For instance, while the extended 70 km link reaches its service cutoff at approximately 8.5\% QBER, the 30 km link sustains valid consensus as the intrinsic error rate approaches 9\%, suggesting that lower channel loss can partially compensate for increased device noise. All results indicate that the short links ($R=30$ km) maintain operational consensus at higher error rates compared to the extended links ($R=70$ km), providing a partial compensation effect between channel loss and device noise. 
This partial compensation effect occurs because the reduced fiber attenuation in shorter links leaves a larger tolerable error budget for intrinsic detector imperfections before breaching the information-theoretic security threshold. Consequently, from an engineering deployment perspective, cost-effective legacy detectors can be safely repurposed for compact metro-core rings, whereas extending the network to regional scales strictly mandates the use of state-of-the-art ultra-low-noise superconducting detectors.

\begin{figure}[t]
    \centering
    \includegraphics[width=0.95\columnwidth]{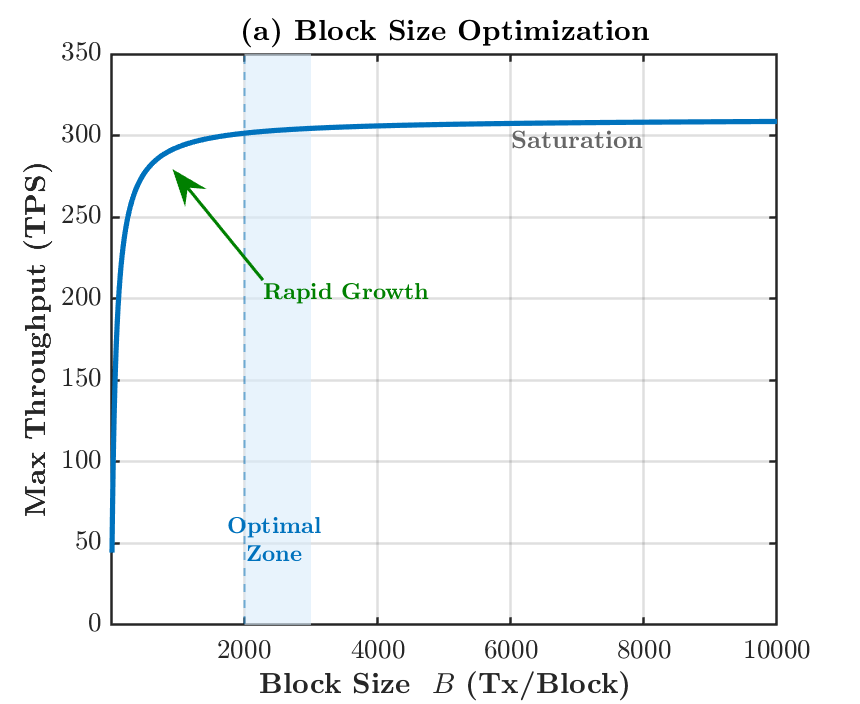}
    \vspace{1em}
    \\
    \includegraphics[width=0.95\columnwidth]{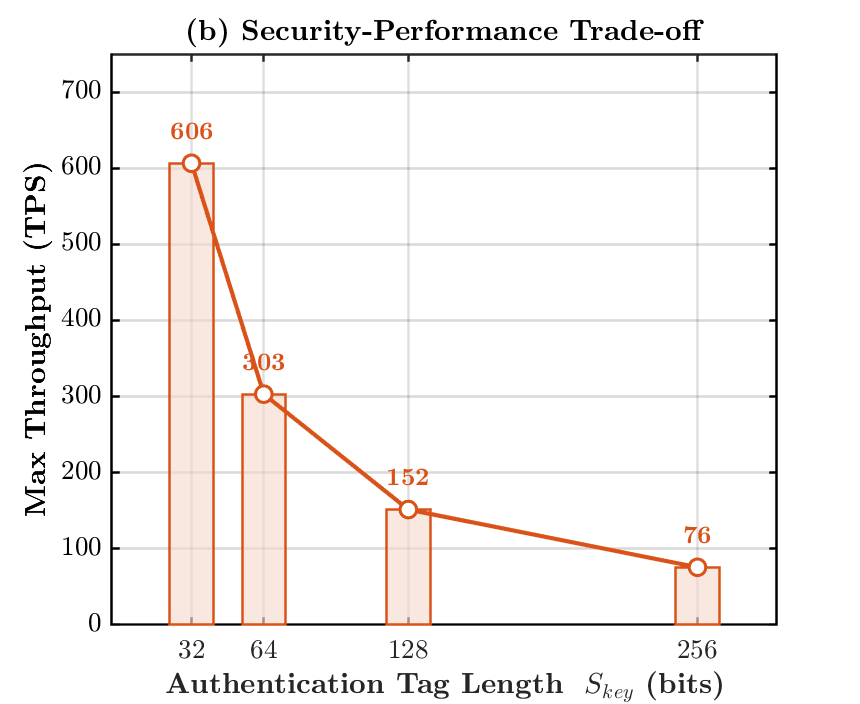}
    
    \caption{{Impact of blockchain protocol parameters on system performance under a metropolitan scenario} (Fixed network parameters: Radius $R=50$ km, Node count $N=20$). (a) Throughput vs. Block Size $B$ ($S_{\text{key}}=64$ bits). (b) Throughput vs. Authentication Tag Length $S_{\text{key}}$ ($B=2500$).}
    \label{fig:protocol_optimization}
\end{figure}

Additionally, following the practical security of the TF-QKD-embedded physical layer, we clarify the other impact of blockchain configurations to identify the suitable operating parameters. Focusing on a typical metropolitan scenario ($R=50$ km, $N=20$), Our finding identify the throughput TPS as a potential parameter while quantifying the throughput under the varied block sizes ($B$) and authentication tag lengths ($S_{\text{key}}$). In Figure \ref{fig:protocol_optimization}(a), it illustrates the effect of block size on the throughput. As $B$ increases, the TPS shows an initial increase driven by the amortization of the consensus key consumption. Nevertheless, the curve tends to saturate beyond $B \approx 2000$, where further increases yield diminishing returns while potentially introducing latency. The saturation of TPS beyond $B=2000$ reflects a marginal utility trade-off. It reveals a certified fact that the efficiency gains from amortizing consensus authentication keys are eventually offset by the increased message propagation latency in the classical overlay. As a result, $B \in [2000, 3000]$ represents the $\it{Sweet~ Spot}$ for this hybrid system, where the cryptographic throughput and network responsiveness achieve the optimal synchronization. In Figure \ref{fig:protocol_optimization}(b), we characterize the trade-off between security and performance. It demonstrates that TPS is inversely proportional to $S_{\text{key}}$. 
The system exhibits a strict inverse proportionality between TPS and $S_{\text{key}}$. Because the WC scheme mandates the strict one-time key usage, doubling the authentication tag length logically consumes double the quantum entropy per transaction, thereby halving the throughput. However, as derived in Appendix~\ref{app:wc_security}, a 64-bit tag already bounds the forgery probability to a negligible $P_{\text{forge}} \le 5.68 \times 10^{-14}$. It is ascertained that $S_{\text{key}}=64$ bits illuminates the optimal cryptographic $\it{Sweet~ Spot}$ for standard consortium applications, while 128-bit tags should be strictly reserved for zero-trust environments where halving the network throughput is a mathematically justifiable trade-off for extreme security.

\subsection{Evaluation and Discussion}

The above performance analysis confirms the implementation feasibility of the proposed framework. All findings highlight the potential role of the TF-QKD-enabled physical layer, which indicates that the quantum-secured blockchain system can sustain the transaction throughput that is suitable for metropolitan-scale consortiums. It can achieve approximately 303 TPS for a network of 20 nodes with a radius of 50 km, Which can be used for supporting connectivity for longer backbone links up to 165 km. Moreover, the robustness analysis suggests that the architecture possesses a degree of resilience against environmental phase noise and hardware degradation, indicating its potential for deployment in the real-world optical fiber environments.

Notably, the current evaluation of the TF-QKD-enabled blockchain may move beyond the traditional paradigm of QKD-based blockchains, which focus performance analysis to the simplified numerical simulations of key generation rates in the physical layer. To the best of our knowledge, this work establishes the rarely mentioned cross-layer global analytical model that quantitatively bridges the quantum key supply ($R_{\text{sup}}$) at the physical layer with the dynamic cryptographic consumption ($K_{\text{dem}}$) driven by the BFT protocol at the logical consensus layer. By explicitly mapping this supply-demand equilibrium, our findings not only validate the practical feasibility of the proposed architecture but also provide a rigorous reference methodology for performance analysis of the future quantum-classical hybrid cryptographic information systems.

\section{Conclusion}
\label{sec:conclusion}

In this paper, we have proposed a quantum-resistant blockchain architecture designed to secure distributed ledgers against the computational threats posed by quantum algorithms, while addressing the scalability challenges observed in traditional QKD-integrated systems. By replacing computationally vulnerable asymmetric cryptography with information-theoretically secure WC authentication and substituting Proof-of-Work with a quantum-secured BFT consensus, this architecture aims to establish a security paradigm that is resilient to both Shor’s and Grover’s algorithms.
An advantage in the quantum blockchain is the exact integration of the TF-QKD protocol within a star-shaped network topology centered on an untrusted relay at just the right station. This design offers a scalable alternative to traditional point-to-point BB84-based schemes. Physically, the TF-QKD protocol overcomes the linear rate-loss limit, enabling secure key distribution over inter-city distances that are usually difficult to achieve without trusted repeaters. Architecturally, the star topology reduces the physical link complexity from quadratic $\mathrm{O}(N^2)$ to linear $\mathrm{O}(N)$, thereby lowering the infrastructure requirements for expanding consortium networks. Notably, the TF-QKD-based physical layer effectively neutralizes detector side-channel risks, maintaining uncompromised system security even with an untrusted central relay. Ultimately, this cross-layer framework bridges the critical gap between theoretical quantum cryptography and applied distributed ledger technology, providing a blueprint for deploying next-generation, unconditionally secure financial infrastructures.

\section*{Acknowledgments}
 This work was supported by the Quantum Science and Technology-National Science and Technology Major Project (Grant No.2021ZD0300700) and the Ye Qixun Science Fund of the National Natural Science Foundation of China (Grant No.U2441219).

\appendices
\section{Derivation of the Asymptotic Secret Key Rate}
\label{app:key_rate_details}

In this appendix, we detail the theoretical framework employed to quantify the asymptotic secret key generation rate ($R_{\text{sup}}$) utilized in the performance analysis. To rigorously evaluate the system's feasibility under realistic conditions, our simulation model integrates the foundational phase-matching TF-QKD protocol established by Lucamarini et al. with the environmental phase noise analysis framework derived by Bertaina et al.

We consider a standard symmetric star-topology network wherein two users, Alice and Bob, transmit optical pulses to a central Untrusted Relay Node (URN). Let $L$ denote the total distance between the users; the fiber length from each user to the URN is thus $L/2$. The channel transmittance $\eta$ for a single arm is modeled as:
\begin{equation}
    \eta = \eta_{\text{det}} \cdot 10^{-\frac{\alpha L}{20}},
\end{equation}
where $\alpha$ is the fiber attenuation coefficient (typically $0.2$ dB/km) and $\eta_{\text{det}}$ accounts for the detection efficiency and internal optical losses. 

In the phase-matching protocol, Alice and Bob encode information into phase-randomized weak coherent pulses and discretize their phase settings into $M$ slices. The URN performs single-photon interference measurements, and valid detection events are retained only when the users' phase slices match. This post-selection mechanism introduces a sifting factor of $1/M$. Furthermore, the finite discretization of the phase space introduces an intrinsic alignment error $E_M$, given by:
\begin{equation}
    E_M = \frac{1}{2} - \frac{\sin(2\pi/M)}{4\pi/M}.
\end{equation}
In our simulation, we set $M=16$ to optimize the trade-off between the sifting efficiency and the intrinsic bit error rate.

To accurately capture the impact of environmental disturbances in field-deployed fibers—such as thermal drift and mechanical vibrations—we model the residual phase instability using the variance $\sigma_\phi^2$. Drawing upon the analysis in Bertaina et al., the additional error contribution induced by phase noise is approximated as \begin{equation}
e_{\text{noise}} \approx \sigma_\phi^2/4.
\end{equation} 
This environmental error is treated as an independent noise source additive to the system's baseline optical misalignment error, $e_{\text{opt}}$. Consequently, the total physical error is described as 
\begin{equation}e_{\text{p}} = e_{\text{opt}} + e_{\text{noise}}.\end{equation} The total phase error rate $e_{\text{ph}}$, governing the privacy amplification process, is derived by probabilistically combining the physical implementation error with the protocol's intrinsic error:
\begin{equation}
    e_{\text{ph}} = e_{\text{p}} + E_M - e_{\text{p}}E_M.
\end{equation}
This formulation allows us to evaluate the system's robustness against varying degrees of environmental noise, ranging from stable laboratory conditions to harsh field environments.

The final secure key rate is calculated using the standard decoy-state method to strictly bound the single-photon contributions. We assume the use of the vacuum+weak decoy-state method with infinite decoy intensities to simulate the asymptotic limit. Let $Q_\mu$ and $E_\mu$ represent the overall gain and quantum bit error rate for signal states with mean photon number $\mu$, respectively. These parameters are derived from the channel transmittance and background dark count rate $Y_0$. The yield of single-photon states, $Y_1$, and the single-photon phase error rate are estimated based on the Poissonian photon number statistics. The asymptotic secret key rate $R$ is explicitly given by:
\begin{equation}
    R = \frac{R_{r}}{M} \left[ \mu e^{-\mu} Y_1 \left(1 - H_2(e_{\text{ph}})\right) - f_{\text{c}} Q_\mu H_2(E_\mu) \right],
\end{equation}
where $R_{r}$ denotes the system repetition rate, $f_{\text{c}}$ is the error correction efficiency factor, and $H_2(x)$ is the binary entropy function given by
\begin{equation}
H_2(x) = -x \log_2 x - (1-x) \log_2 (1-x).
\end{equation}
This analytical model incorporates the quadratic scaling of phase noise ($\sigma_\phi^2$) and the protocol-specific sifting penalty, providing a conservative and physically rigorous estimation of the system's performance boundaries.

\section{Formal Specification of Operational Workflow}
\label{app:protocol_workflow}

\vspace{1em}
\noindent\rule{\linewidth}{0.8pt} \\[4pt]
\textbf{Algorithm 1} Hybrid Quantum Blockchain \\[4pt]
\rule{\linewidth}{0.4pt}

\label{alg:quantum_blockchain_workflow}
\begin{algorithmic}[1]
\small 
\Require Terminal Nodes $\mathcal{N} = \{\mathrm{TN}_1, \dots, \mathrm{TN}_N\}$, max faulty nodes $f$.
\Statex
\Statex \textbf{\textit{Phase 1: Key Establishment (Async)}}
\ForAll{node pair $(\mathrm{TN}_i, \mathrm{TN}_j) \in \mathcal{N} \times \mathcal{N}$}
    \State Generate ITS keys via TF-QKD \& central URN
    \State Stratify keys: Evidence ($K_{\text{evid}}$), Consensus ($K_{\text{cons}}$)
    \State Securely store keys in local reserves
\EndFor
\Statex
\Statex \textbf{\textit{Phase 2: Transaction Submission}}
\Procedure{SubmitTx}{$\mathrm{TN}_i, M$}
    \State $\text{Ct}_i \gets \text{Ct}_i + 1$ \Comment{Local nonce}
    \State Fetch key pair $(k_{\text{hash}}, k_{\text{otp}}) \in K_{\text{evid}}$ via index $\text{Ct}_i$
    \ForAll{$j \in \{1, \dots, N\}$}
        \If{$i \neq j$}
            \State $\tau_{i,j} \gets h_{k_{\text{hash}}}(M \parallel \text{Ct}_i) \oplus k_{\text{otp}}$
        \Else
            \State $\tau_{i,j} \gets \mathbf{0}$
        \EndIf
    \EndFor
    \State Broadcast $Tx_{\text{req}} = \langle ID_i, \text{Ct}_i, M, \mathcal{V}_i \rangle$ to Leader
\EndProcedure
\Statex
\Statex \textbf{\textit{Phase 3: Consensus and Verification}}
\Procedure{VerifyAndVote}{$\mathrm{TN}_j, B_{\text{cand}}$}
    \ForAll{$Tx_{\text{req}} \in B_{\text{cand}}$ from sender $\mathrm{TN}_s$}
        \State Fetch local $(k_{\text{hash}}, k_{\text{otp}})$ via index $\text{idx}_{sj}$
        \State $\tau'_{s,j} \gets h_{k_{\text{hash}}}(\text{idx}_{sj} \parallel M) \oplus k_{\text{otp}}$
        \If{$\tau'_{s,j} \neq \tau_{s,j}$ \textbf{or} $\neg \Call{Valid}{M}$}
            \State \Return \textbf{Reject} $B_{\text{cand}}$
        \EndIf
    \EndFor
    \State Broadcast \texttt{ACCEPT} vote using $K_{\text{cons}}$
    \If{Valid \texttt{ACCEPT} votes $\ge 2f+1$}
        \State \Call{FinalizeAndReveal}{$B_{\text{cand}}$}
    \EndIf
\EndProcedure
\Statex
\Statex \textbf{\textit{Phase 4: Finalization and Disclosure}}
\Procedure{FinalizeAndReveal}{$B_N$}
    \State Finalize $B_N$ and compute header hash $H(B_N)$
    \ForAll{Sender $\mathrm{TN}_i$ with valid $Tx \in B_N$}
        \State Disclose used OTP keys via classical channel
    \EndFor
    \State Next Leader embeds keys into block $B_{N+1}$
\EndProcedure
\end{algorithmic}
\vspace{-0.5em}
\noindent\rule{\linewidth}{0.8pt}
\vspace{1em}

\section{Security Bound of the Wegman-Carter Authentication}
\label{app:wc_security}

In our proposed architecture, the integrity of transactions and consensus messages is enforced via the Wegman-Carter authentication scheme. Unlike computational signatures such as ECDSA, which depend on unproven hardness assumptions, the WC scheme provides unconditional security resilience against quantum adversaries.

Let $\mathcal{H}$ denote a family of $\epsilon$-Almost XOR Universal ($\epsilon$-AXU) hash functions mapping a message space $\mathcal{M}$ to a tag space $\mathcal{T} = \{0, 1\}^{S_{\text{key}}}$, where $S_{\text{key}}$ represents the bit length of the authentication tag. For an arbitrary message $m \in \mathcal{M}$, the authentication tag $\tau$ is computed as:
\begin{equation}
    \tau = h_k(m) \oplus k_{\text{otp}}.
\end{equation}
This construction relies on two distinct secret elements retrieved from the TF-QKD key pool: a secret hash key $k$, which selects a specific function $h_k \in \mathcal{H}$, and a fresh OTP key $k_{\text{otp}} \in \{0, 1\}^{S_{\text{key}}}$. A critical constraint of this protocol is that $k_{\text{otp}}$ must be unique for every message and never reused, ensuring the scheme's information-theoretic properties.

The security of this mechanism is anchored in the perfect secrecy of the OTP. Given that $k_{\text{otp}}$ is uniformly random and unknown to the adversary, the resulting tag $\tau$ is statistically independent of the hash output $h_k(m)$. Consequently, the observation of $\tau$ yields no information regarding $h_k(m)$ or the hash key $k$, regardless of the adversary's computational resources. Even a quantum adversary possessing infinite computing power cannot invert the OTP encryption to analyze the underlying hash function structure.

Under these conditions, the adversary's optimal strategy is limited to a blind forgery attempt, seeking to generate a valid tag $\tau'$ for a modified message $m' \neq m$. The probability of a successful forgery is strictly bounded by the combinatorial properties of the $\epsilon$-AXU hash family rather than computational hardness. This probability is derived as:
\begin{align}
    P_{\text{forge}} &= \max_{m \neq m', \tau, \tau'} \Pr_k \big[ h_k(m') \oplus k_{\text{otp}} = \tau' \nonumber \\
              &\qquad\qquad\quad \mid h_k(m) \oplus k_{\text{otp}} = \tau \big] \nonumber \\
              &= \max_{m \neq m', \delta} \Pr_k \big[ h_k(m) \oplus h_k(m') = \delta \big] \nonumber \\
              &\leq \epsilon, \label{eq:p_forge}
    \end{align}
where $\delta = \tau \oplus \tau'$ denotes the differential in the tag space.

Adopting a standard polynomial hash function over a Galois Field $GF(2^{S_{\text{key}}})$, the collision probability $\epsilon$ is determined by the ratio of the message length $L$ to the tag space size, approximated as:
\begin{equation}
    \epsilon \approx \frac{L}{2^{S_{\text{key}}}}.
\end{equation}
Within the context of our simulation parameters, we let $S_{\text{key}} = 64$ bits and the maximum block length is $L = 2^{20}$, and then the upper bound for a successful forgery is calculated as:
\begin{equation}
    P_{\text{forge}} \le \frac{2^{20}}{2^{64}} = 2^{-44} \approx 5.68 \times 10^{-14}.
\end{equation}
This negligible probability confirms that the system maintains robust information-theoretic security. By consuming only 64 bits of fresh QKD key material per message, the architecture effectively eliminates vulnerabilities associated with quantum computing attacks on the authentication layer.

\bibliographystyle{IEEEtran}
\bibliography{mybibfile} 
\end{document}